\documentclass[a4paper,11pt]{article}
\pdfoutput=1 

\usepackage{jinstpub} 
\usepackage{amsmath}

\title{\boldmath Beam dynamics challenges in linear colliders based on laser-plasma accelerators}

\author[a,b]{C. B. Schroeder,}
\author[a]{C. Benedetti,}
\author[a]{S. S. Bulanov,}
\author[a]{D. Terzani,}
\author[a]{E. Esarey,}
\author[a]{and C. G. R. Geddes}

\affiliation[a]{Lawrence Berkeley National Laboratory, Berkeley, California 94720 USA}
\affiliation[b]{Department of Nuclear Engineering, University of California, Berkeley, California 94720 USA}

\emailAdd{cbschroeder@lbl.gov}

\abstract{In this paper we discuss design considerations and beam dynamics challenges associated with laser-driven plasma-based accelerators as applied to multi-TeV-scale linear colliders.
Plasma accelerators provide ultra-high gradients and ultra-short bunches, offering the potential for compact linacs and reduced power requirements.  We show that stable, efficient acceleration with beam quality preservation is possible in the nonlinear bubble regime of laser-plasma accelerators using beam shaping.  Ion motion, naturally occuring for dense beams (i.e., low emittance and high energy) severely damps transverse beam instabilities.  Coulomb scattering by the background ions is considered and it is shown that the strong focusing in the plasma strongly suppresses scattering-induced emittance growth. Betatron radiation emission from the transverse motion of the beam in the plasma will result in beam power loss and energy spread growth; however for sub-100 nm emittances, the beam power loss and energy spread growth will be sub-percent for multi-TeV-class plasma linacs. }


\proceeding{ICFA Beam Dynamics Newsletter}

\begin{document}
\maketitle
\flushbottom

\section{Introduction}
\label{sec:intro}

The discovery potential of a collider is determined by the energy reach and integrated luminosity.   The energy reach of a linear collider is limited by the accelerating gradient of the accelerator technology and the luminosity is limited by the power costs.  For example, the proposed Compact Linear Collider (CLIC) relies on conventional metallic RF structures with a peak accelerating gradient of 100~MV/m, and the size of the linac to reach 3~TeV center-of-mass energy is 50~kilometers, achieving luminosities of $6\times 10^{34}$~cm$^{-2}$s$^{-1}$ with an estimated power consumption above 500~MW \cite{CLIC-CDR}.  Laser-driven plasma-based accelerators (LPAs) \cite{Esarey09,Hooker13} have demonstrated gradients 1--100~GV/m \cite{Gonsalves19}, orders of magnitude larger than conventional RF accelerators.  A linear collider based on LPAs offers the possibility of orders of magnitude reduction in the size of the collider linacs, and the expected associated reductions in cost. In addition, LPAs naturally accelerate ultra-short bunches (sub-100~fs), significantly reducing the beamstrahlung during the collision, and, hence, reducing the required beam power to reach a luminosity goal.

In this article, we discuss some of the basic design considerations for a linear collider based on laser-plasma accelerators and some of the unique beam dynamics challenges in plasma accelerators.  Although the focus in this article is laser-driven plasma acceleration, many of the beam dynamics challenges described below are common to both beam-driven and laser-driven plasma-based accelerators.  In particular, we address the issues of witness beam transverse stability, scattering by background ions, and energy loss via betatron radiation.  It is shown that these effects do not limit the performance of plasma accelerators up to beam energies of $\sim$10~TeV.

\section{Considerations for laser-plasma-based collider design}

A conceptual design of an LPA-based collider has not been completed. However, several preliminary studies have been performed \cite{Schroeder10b,Schroeder16}. These studies have been used to help identify the expected operational plasma and laser parameters to guide the R\&D toward developing plasma-based collider concepts \cite{ANAR17}.  Many of the parameters are determined by the plasma density choice, and the density regime considered in these studies, operating at a plasma density of the order of $n\sim 10^{17}$~cm$^{-3}$, provides for a large accelerating gradient and reduces the linac power requirements, while keeping the beamstrahlung at the collision point at an acceptable level for a target luminosity. 

\subsection{General linear lepton collider considerations}
\label{sec:general}

The luminosity determines the rate of a physical process given its cross section.   In general, the cross section decreases quadratically with energy ($\propto 1/\gamma^2$), and, hence, as the energy of the collider increases, ideally the luminosity should be increased quadratically.  
For symmetric electron and positron beams at the interaction point (IP), the luminosity is
\begin{equation}
\mathcal{L} = \frac{f N^2}{4 \pi \sigma_x^* \sigma_y^*} =  \frac{P_b}{U_b} \frac{N}{4 \pi \sigma_x^* \sigma_y^*}  =  \frac{\eta P}{\sqrt s} \frac{N}{4 \pi \sigma_x^* 
\sigma_y^*}
,
\end{equation}
where $N$ is the number of particles per bunch, $f$ the repetition rate, $U_b = \gamma m c^2$ the beam energy,
$\sqrt s = 2U_b$ is the center-of-mass energy, $P_b= f N U_b $ is the beam power, and $P = 2P_b/\eta$ is the wall-plug power,  with $\eta$ the overall efficiency from wall to beam.  The rms beam sizes at the IP are determined by the normalized horizontal and vertical rms transverse emittances $\epsilon_{x,y}$ and final focus beta-functions $\beta_{x,y}^*$, $\sigma_{x,y}^* = (\beta_{x,y}^* \epsilon_{x,y}/\gamma)^{1/2}$.

An important figure of merit is the luminosity per wall-plug power $\mathcal{L}/P$.  Reducing the beam cross section at the IP, $\sigma_x^* \sigma_y^*$, will reduce power costs to achieve a desired luminosity.  However this is limited by the achievable emittance and the final-focus optics.  The emittance will be limited by what is achievable at the injector, the beam cooling, and emittance growth in the plasma linacs.  The final-focus optics are ultimately limited by the Oide effect \cite{Oide88}, i.e., strong focusing leads to synchrotron radiation, and the resulting chromatic effects increase the beam size.  For example, if we consider a flat beam (to reduce beamstrahlung, as discussed below), then the spot size at the final focus in the minor (vertical) dimension including the Oide effect, is
\begin{equation}
\sigma_{y}^{*2} =  \beta_y^{*} \epsilon_y/\gamma + \frac{110}{3 \sqrt{6\pi}} \frac{r_e^2}{\alpha}  \left( \frac{\gamma \epsilon_y }{ \beta_y^*} \right)^{5/2} F
,
 \end{equation}
where $r_e = e^2/mc^2$ is the classical electron radius (formulas are given in Gaussian units throughout the paper), $\alpha = e^2/\hbar c$  is the fine structure constant, and $F$ is a dimensionless function of the specific focusing system configuration \cite{Oide88}.  At high energies and small final focus beta functions, the beam size will be limited by the Oide effect, and there is an optimal $\beta_y^*$ to achieve the minimum beam size $(\sigma_{y}^*)_{\rm min} \propto F^{1/7}\epsilon_y^{5/7}$.

Increasing the number of particles per bunch $N$ also increases the luminosity per wall-plug power $\mathcal{L}/P$.  However, the bunch charge will be limited by beamstrahlung at the IP, radiation emitted during the beam-beam collision when one beam experiences the fields of the colliding beam.  This effect results in an increased background from the emitted radiation and energy loss in the beam, inducing a large beam energy spread. 
The beam-beam interaction is characterized by the Lorentz-invariant beamstrahlung parameter (mean field strength in the beam rest frame
normalized to the Schwinger critical field $E_c = m^2c^3/e\hbar$): $\Upsilon = \gamma \langle E + B \rangle /E_c$.  For a Gaussian beam the average beamstrahlung parameter is \cite{YokoyaChen92}
\begin{equation}
\Upsilon \approx \frac{5r_e^2}{6 \alpha} \frac{\gamma}{\sigma_z}  \frac{N}{(\sigma_x^* + \sigma_y^*)},
\label{eq:Y}
 \end{equation}
 where $\sigma_z$ is the rms bunch length.
 The average number of beamstrahlung photons emitted per electron at the IP is \cite{YokoyaChen92}
 \begin{equation}
n_\gamma \approx 2.5 \left( \frac{\alpha^2}{ r_e} \frac{\sigma_z}{\gamma}  \right)\frac{\Upsilon}{(1+ \Upsilon^{2/3})^{1/2}} .
\label{eq:n_g}
 \end{equation}
Associated with this radiation emission is a relative beam energy spread increase \cite{YokoyaChen92}
 \begin{equation}
\delta_\gamma \approx 1.2 \left( \frac{\alpha^2}{ r_e} \frac{\sigma_z}{\gamma}  \right)\frac{\Upsilon^2}{\left[1+ (3\Upsilon/2)^{2/3}\right]^{2}} ,
\label{eq:d_g}
 \end{equation}
 and, hence, broadening of the luminosity energy spectrum.  
 As eq.~\eqref{eq:Y} indicates, flat beams $\sigma_y^* \ll \sigma_x^*$ at the IP can reduce the beamstrahlung effect, while preserving high luminosity.   
Conventional linear collider proposals typically operate in the regime $\Upsilon \ll 1$, or classical beamstrahlung regime.   In the classical beamstrahlung regime, 
 \begin{equation}
\frac{\mathcal{L}}{P} \approx 0.04 \left( \frac{\eta n_\gamma}{\alpha r_e\sqrt s} \right) \left( \frac{\sigma_x^* + \sigma_y^*}{\sigma_x^*\sigma_y^*} \right) .
 \end{equation}
As eq.~\eqref{eq:Y} indicates the beamstrahlung parameter increases with increasing energy and shorter bunches ${\Upsilon}\propto \gamma/\sigma_z$.  
Plasma accelerators are best suited to accelerate ultra-short beams, a fraction of the plasma wavelength, or $\sigma_z 
\sim 10~\mu$m operating at $n\sim10^{17}$~cm$^{-3}$.
Hence plasma-based colliders operating at high energy ($U_b\gtrsim 1$~TeV) will be in the high-field beamstrahlung regime, or quantum beamstrahlung regime, such that $\Upsilon \gg 1$.  
In the high-field beamstrahlung regime
 \begin{equation}
\frac{\mathcal{L}}{P} \approx 0.02 \left(\frac{\eta n_\gamma^{3/2 } }{\alpha^2 \sqrt{r_e} \sqrt s}  \right)
\left( \frac{\sigma_x^* + \sigma_y^*}{\sigma_x^*\sigma_y^*} \right) \frac{\gamma^{1/2}}{\sigma_z^{1/2}} .
 \end{equation}
In this regime, short beams increase the luminosity per power \cite{Dugan94}.  

Assuming the luminosity must increase quadratically with center-of-mass energy and there is the constraint that there is a fixed beamstrahlung photons per lepton $n_\gamma$ that the detector can tolerate, the linear collider wall-plug power required, in the high-field beamstrahlung regime using flat beams, scales as 
 \begin{equation}
P \propto  (\sqrt s)^{5/2} \epsilon_y^{1/2} \sigma_z^{1/2} . 
\label{eq:p-scaling}
 \end{equation}
As the center-of-mass energy increases, the wall-plug power to achieve the required luminosity increases as in eq.~\eqref{eq:p-scaling}.  The ability to generate low emittance beams and shorter beams saves power.  Plasma-based accelerators are, hence, of interest for linear collider applications because of the ultra-high accelerating gradients and also because plasma-based accelerators naturally accelerate short beams, with bunch lengths that are a fraction of plasma wavelength.  For typical plasma densities, the bunch length is on the order of ten microns, orders of magnitude shorter than conventional accelerators, potentially yielding significant power savings. 
Note that the constraints owing to beamstrahlung are present for $e^{-}e^{+}$ or $e^{-}e^{-}$ collisions, and one may consider a $\gamma\gamma$ collider (or $\gamma e^{-}$ collisions)  to remove the beamstrahlung constraints.

\subsection{Basic laser-plasma-accelerator-based linac design}
\label{sec:LPA}

In this section, we discuss the basics of LPAs before addressing some beam dynamics issues unique to plasma acceleration in section~\ref{sec:beam-dyn}.
In an LPA a large amplitude electron plasma wave is excited by the ponderomotive force (radiation pressure) of a laser with relativistic intensity such that $a^2 \sim 1$, with  $a^2  = 0.73 (\lambda [\mu {\rm m}])^2 I_L[10^{18} \textrm{W/cm}^2]$, where $I_L$ is the laser intensity and $\lambda$ the laser wavelength.  In the standard configuration, often referred to as a laser wakefield accelerator (LWFA), a single laser pulse with duration resonant with the plasma wavelength $\tau_L \sim \omega_p^{-1}$ is used to resonantly drive the plasma wave with relativistic phase velocity, near the group velocity of the laser propagating in an underdense plasma.   Here $\omega_p = k_p c= (4\pi n e^2/m)^{1/2}$ is the electron plasma frequency.   Relativistic charged particles co-propagating behind the laser at an appropriate phase in the laser-excited plasma wave may be accelerated to high energy. 

There are several regimes of laser-driven plasma acceleration that may be accessed based on the intensity of the laser pulse.  Collider designs based on operation in the quasi-linear regime with $(a^2/2)(1+a^2/2)^{-1/2} \lesssim 1$ have been explored \cite{Schroeder16}.    For high laser intensities $a^2\gg1$, the LPA can operate in the bubble regime, where (almost) all the electrons are expelled by the laser ponderomotive force, forming an ion cavity co-propagating behind the laser.  Balancing the laser ponderomotive force with the space-charge force of the ion cavity yields the condition  to form an ion cavity, $2 \sqrt{a} \gtrsim k_p r_L$.   The plasma wakefield structure inside the ion cavity in the bubble regime is 
\begin{align}
\label{eq:ez}
&E_z/E_0 = k_p \zeta /2 ,
\\
\label{eq:er}
&(E_r-B_\theta)/E_0 = k_p r /2 , 
\end{align}
with $E_0 = mc^2k_p/e$, or $E_0 [\textrm{V/m}] \approx 96 (n[\textrm{cm}^{-3}])^{1/2}$ the characteristic size of the wakefields.  Here $\zeta = z-ct$ is the co-moving longitudinal coordinate ($\zeta=0$ corresponds to the center of the ion cavity) and $r$ is the radial coordinate.  For plasma densities $n\sim 10^{18}~\textrm{cm}^{-3}$ the characteristic accelerating field is $E_0 \sim 100$~GV/m (i.e., three orders of magnitude higher than CLIC). 
In the bubble regime, the accelerating field $E_z$ is independent of the transverse position and the focusing field $E_r-B_\theta$ is linear with respect to the transverse coordinate and independent of the axial position (conserving the electron beam transverse normalized rms emittance).  
Note that the transverse fields in the ion cavity are defocusing for positrons; hence, stable positron acceleration can not occur in the bubble regime.  For stable positron acceleration, the plasma accelerator must operate in the quasi-linear regime or the wakefield structure must be modified.  Wakefield excitation in plasma columns have been proposed for modifying the wakefield to allow for positron focusing and acceleration \cite{Diederichs19}.   

\begin{figure}[thbp]
\centering 
\includegraphics[width=1\textwidth]{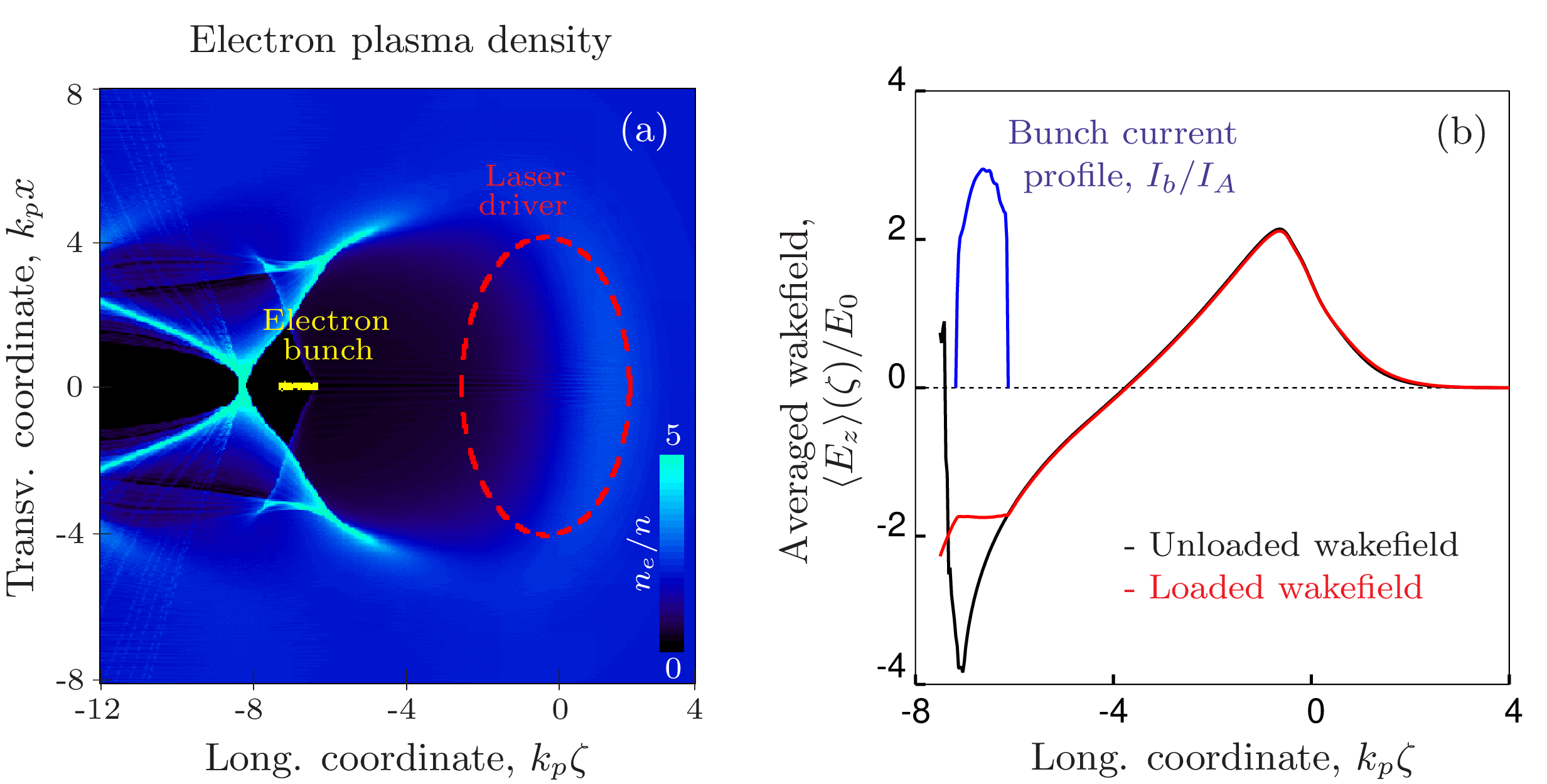}
\caption{\label{fig:LPA-stage}
Two dimensional $(\zeta, x)$ map (a) of the electron plasma density for an LPA stage in the bubble regime. The parameters of the bi-Gaussian laser driver (red dashed ellipse) are $a_0=4.5$, $\omega_p\tau_L=2.7$ (FWHM duration of the intensity), and $k_p r_L =4$. The accelerating electron bunch is shown in yellow. Lineout (b) of the on-axis longitudinal wakefield averaged over the accelerator length for a loaded (red line) and an unloaded (black line) stage. The optimal bunch current profile that produces the constant accelerating field along the bunch is shown in blue.}
\end{figure}

Figure~\ref{fig:LPA-stage} (a) shows an example of the ion cavity formed by an intense laser in the nonlinear regime. In the bubble regime, the laser effectively creates a plasma channel and can self-guide over a distance corresponding to many Rayleigh ranges.  The condition for matched guiding in the bubble regime is $k_p r_L \approx 2 \sqrt{a}$ \cite{Lu07}, with the laser power normalized to the critical power $P/P_c = a^3/8$.  With matched guiding, the bubble radius is approximately $R \approx r_L \approx 2 \sqrt{a}/k_p$ \cite{Benedetti13}.

An LPA-based collider would consist of multiple LPA accelerating stages.   Figure~\ref{fig:LPA-stage} shows an example of a single LPA stage modeled with the Particle-In-Cell code {INF\&RNO} \cite{INFERNO1, INFERNO2}. In the simulation the wakefield and the laser are modeled assuming a 2D axisymmetric geometry that correctly reproduces the 3D field structure and laser evolution, while the beam particles are pushed in 3D. Here the laser and plasma parameters are $a=4.5$, $\omega_p\tau_L=2.7$ (FWHM duration of the intensity), $k_p r_L =4$, and $n=4.6\times 10^{17}~\textrm{cm}^{-3}$. The corresponding laser energy is $U_L=50$~J.  This example accelerates a matched 1.2~nC, 2.2~$\mu$m rms length bunch with an average loaded gradient of 117~GV/m (or $E_z/E_0 = 1.8$), achieving 3.2~GeV energy gain per stage with a stage length $L_s=$3.1~cm.  By properly shaping the electron beam current, the longitudinal dependence of the beam-loaded accelerating field can be controlled, minimizing energy spread growth. %
Note that a linear density taper is used to control bunch-laser dephasing, with a final density after 3.1~cm of $8\times 10^{17}~\textrm{cm}^{-3}$. The laser energy depletion at the end of the stage is 20\%. (In principle, the majority of the remaining laser energy could be recovered with a photovoltaic.)  The wake to beam energy efficiency of this example is 43\%.

Many stages would be used to reach TeV energies (for this stage example, 218 stages to reach 1~TeV beam energy).  
Note that plasma mirrors can allow for compact coupling of lasers between stages, yielding 
geometric/average accelerating gradients of a few GV/m.  Initial proof-of-principle staging
experiments using plasma mirrors have been performed \cite{steinke16}.   Methods for coupling beams between LPA stages, with high charge
capture and preservation of beam quality must be developed further, in addition to beyond state-of-the-art
beam and laser alignment methods relying on kHz feedback.  
Energy recovery may be employed to improve the overall efficiency, including recovery of energy remaining in the plasma wave and the depleted laser driver \cite{Schroeder16}.

Many of the important parameters for LPA design scale with the plasma density and laser wavelength. With the density and wavelength scalings known, a single design may be scaled to explore parameter space, given additional constraints and for different laser driver technologies. For fixed collider parameters (center-of-mass energy, luminosity, and IP focusing) and fixed normalized
LPA parameters (normalized laser intensity $a_0$, laser transverse size $ k_pr_L$, and laser pulse duration $\omega_p \tau_L$), the basic scalings \cite{Schroeder10b} with plasma density and laser wavelength  are given in Table \ref{tab:scalings}.

\begin{table}[htbp]
\centering
\caption{\label{tab:scalings} Basic LPA and collider scalings with plasma density and laser wavelength.}
\smallskip
\begin{tabular}{l  c}
\hline
Accelerating field  & $E_z \propto n^{1/2}$ \\
LPA stage length  & $L_s \propto n^{-3/2}\lambda^{-2}$ \\
Energy gain per stage & $\Delta \gamma \propto n^{-1}\lambda^{-2}$  \\
Number of stages  & $N_s \propto n \lambda^2$ \\
Bunch charge & $N_b \propto n^{-1/2}$ \\
Bunch length & $L_b \propto n^{-1/2} $ \\
Laser energy & $U_L \propto n^{-3/2} \lambda^{-2}$ \\
Pulse duration & $\tau \propto n^{-1/2}$ \\
Repetition rate & $f \propto n$ \\
Beamstrahlung photons/$e$ & $n_\gamma \propto n^{-1/2}$ \\
Wall-plug power & $P \propto n^{1/2}$ \\
\hline
\end{tabular}
\end{table}

The scaling with wall-plug power implies operation at lower densities for reduced power costs $P\propto n^{1/2}$.   These power savings are the result of a higher bunch charge at lower density $N \propto n^{-1/2}$.  However, as the plasma density is decreased, the gradient is reduced $E_z \propto n^{1/2}$, and the required laser energy per pulse $U_L \propto n^{-3/2} \lambda^{-2}$ and average laser power $P_{L} = f U_L \propto n^{-1/2} \lambda^{-2}$ increase.   The charge per bunch at the IP will be constrained by beam-beam effects, as discussed in section~\ref{sec:general}, and the number of beamstrahlung photons (and the associated beamstrahlung induced beam energy spread) will limit the charge per bunch.  Given these collider constraints plasma densities in the range $10^{16}$--$10^{19}$~cm$^{-3}$ may be considered: below a few $10^{16}$~cm$^{-3}$ the gradient becomes too low and the beamstrahlung effects too severe, owing to the increased charge per bunch, and above a few $10^{18}$~cm$^{-3}$ the power required to achieve a target luminosity becomes too large.  Also note that the wall-plug power is independent of laser wavelength.  This implies that different laser technologies and amplifying media may be considered for a laser-plasma-based collider.  Two potential technologies that could achieve the required peak and average laser power are coherent combination of fiber lasers \cite{Zhou15,Stark21}, operating at $\lambda \simeq 1~\mu$m, and lasers based on Tm:YLF \cite{Tamer21}, operating at $\lambda \simeq 1.9~\mu$m.

\section{Beam dynamics challenges in plasma accelerators}
\label{sec:beam-dyn}

In this section we address three of the main beam dynamics challenges that are unique to plasma-based accelerators.  In section \ref{sec:hose} we consider the transverse stability of the accelerated electron beam in the bubble regime.  Ion motion, induced by the high density bunches, provides a natural method to stabilize the beam.  In section \ref{sec:scatt} we discuss emittance growth from scattering with charged particles in the background plasma.  The strong focusing in the plasma prevents significant emittance growth from the Coulomb scattering.  Lastly, in section \ref{sec:rad}  we discuss the synchrotron radiation generated by the transverse motion of the beam in the plasma.  The induced energy spread from the betatron radiation is proportional to the beam emittance, and for collider-relevant emittances the induced energy spread from betatron radiation is a fraction of a percent.  These results indicate that total required collider power will likely be the limiting factor for very high energy linear colliders $\sqrt{s}>$15~TeV.

\subsection{Transverse stability}
\label{sec:hose}

Transverse beam stability, i.e., suppressing beam hosing \cite{Whittum91, Huang07, Mehrling18b}, has been identified as a critical challenge toward realizing a plasma-based collider. In beam hosing, the excited transverse wakefield of a beam couples to the beam transverse position, leading to exponential growth in the beam centroid displacement. This implies that small asymmetries or misalignments are exponentially amplified during the acceleration process, potentially leading to a strong degradation of the beam emittance, or beam loss.  
Plasma accelerators operate in a strongly-beamloaded regime for high efficiency.  In this regime, the wakefields generated by the accelerated beam are of the order of the ion cavity fields $\sim E_0$, orders of magnitude stronger than in conventional accelerators.  These large transverse wakefields will drive the instability, and without transverse instability mitigation, beam hosing will limit the achievable efficiency \cite{Lebedev17}.

Variation of the transverse wakefields along the beam (head-to-tail) can mitigate hosing \cite{YYLau89,Lehe17}. This mechanism is similar to the Balakin-Novokhatsky-Smirnov (BNS) damping of the beam-breakup instability in conventional accelerators.  In the nonlinear bubble regime, the betatron wavenumber is $k_\beta = k_p /\sqrt{2\gamma}$.  Hence longitudinal variation in focusing force may be provided by a beam energy chirp $k_\beta(\zeta) = k_p /\sqrt{2\gamma(\zeta)}$.  However, in the strongly-beamloaded, high-efficency regime, the required energy variation is  high and may not be compatible with beam transport between stages and the collider beam delivery system.  Stability may also be achieved by varying the background ion density along the beam $k_\beta(\zeta) = k_p(\zeta) /\sqrt{2\gamma}$.
For sufficiently dense beams, ion motion will occur on an electron plasma period \cite{Rosenzweig05, Benedetti17, An17}.  This ion motion will naturally generate a head-to-tail variation in the focusing fields provided by the background ions, and this longitudinal variation in focusing results in a BNS-type damping of the hosing instability \cite{Mehrling18b}. Hence, ion motion can allow for stable acceleration of witness beams in plasma-based accelerators without energy spread. 

For a matched beam in the focusing field provided by the (unperturbed, uniform) background ions, the rms size of the beam is 
\begin{equation}
\sigma_x = \left( \frac{2 \epsilon_x^2}{k_p^2 \gamma} \right)^{1/4} .
\end{equation}
As the beam accelerates, the transverse size of the beam decreases adiabatically, increasing the beam density $n_b = n_{bi} (\gamma/\gamma_i)^{1/2}$, where $n_{bi}$ is the bunch density at injection.  For a sufficiently high beam current, the space charge force of the beam will move the ions on the time scale of the beam duration, and the condition such that ion motion occurs may be expressed as $\Lambda \sim 1$ \cite{Benedetti17}, with 
\begin{equation}
\Lambda = Z_i \frac{m}{M_i} \frac{I_b}{I_A} \frac{L_b^2}{\sigma_x^2}  \approx  
0.2\frac{Z_i}{A_i} \frac{I_b[kA]  (L_b[\mu \textrm{m}])^2}{\epsilon_x [\textrm{nm}]} \left( {U_b[\textrm{GeV}] n[10^{18}~\textrm{cm}^{-3}]}\right)^{1/2} ,
\label{eq:ion-motion}
\end{equation}
where $Z_i$ is the charge state of the ions, $M_i \approx A_i m_p$ is the mass of the ions, $I_b$ is the beam current, $I_A = mc^3/e$ is the Alfv\'{e}n current, and $L_b$ is the bunch length (a flat-top profile is assumed).  A beam with collider-relevant emittance will be in a regime with $\Lambda \sim 1$ at relatively low energy.  For example $\Lambda \sim 1$ for a 3~GeV, 1~kA beam, with $L_b = 10~\mu$m, $\epsilon_x = 10$~nm, propagating in a $10^{17}~\textrm{cm}^{-3}$ plasma.  Hence, ion motion will be present at the first stage of a plasma accelerator, and $\Lambda \gtrsim 1$ for subsequent stages.

Ion motion will modify the transverse wakefield provided by the background ions, inducing a longitudinal head-to-tail variation in the focusing field.  For $\Lambda <1$, the transverse wakefield has the form \cite{Benedetti17}
\begin{equation}
(E_r-B_\theta)/E_0 = \frac{k_p r}{2} \left[ 1 + \Lambda \frac{\zeta^2}{L_b^2} H(r) \right] ,
\label{eq:wake-ion}
\end{equation}
where $H(r) \sim 1$ is a radial function determined by the transverse beam profile. For a Gaussian transverse beam profile, $H(r) = [1- \exp (-q)]/q$, with $q=r^2/2\sigma_x^2$. Equation~\eqref{eq:wake-ion} assumes a longitudinally uniform bunch current profile, with the bunch head located in $\zeta=0$.  The focusing field has a quadratic dependence from head to tail and this results in decoherence and suppression of the hosing
(beam centroid) growth.  The decoherence length may be approximated as $L_{d} \sim 4\pi/(\Lambda k_\beta)$, where $k_\beta$ is the betatron wavenumber.  For $\Lambda > 1$, the beam is stabilized with respect to hosing within a fraction of a betatron wavelength \cite{Mehrling18b}.
Note that the ion motion-induced perturbation of the longitudinal wakefield is generally negligible. This can be seen using the
Panofsky–Wenzel theorem $\partial_\zeta (E_r-B_\theta) = \partial_r E_z$, yielding the longitudinal wake perturbation at the beam tail $ \vert \delta E_z/E_0\vert \sim Z_i (m/M_i) k_p r_e N$, and $\vert \delta E_z/E_0\vert \ll 1$ for typical parameters.

\begin{figure}[htbp]
\centering 
\includegraphics[width=0.6\textwidth]{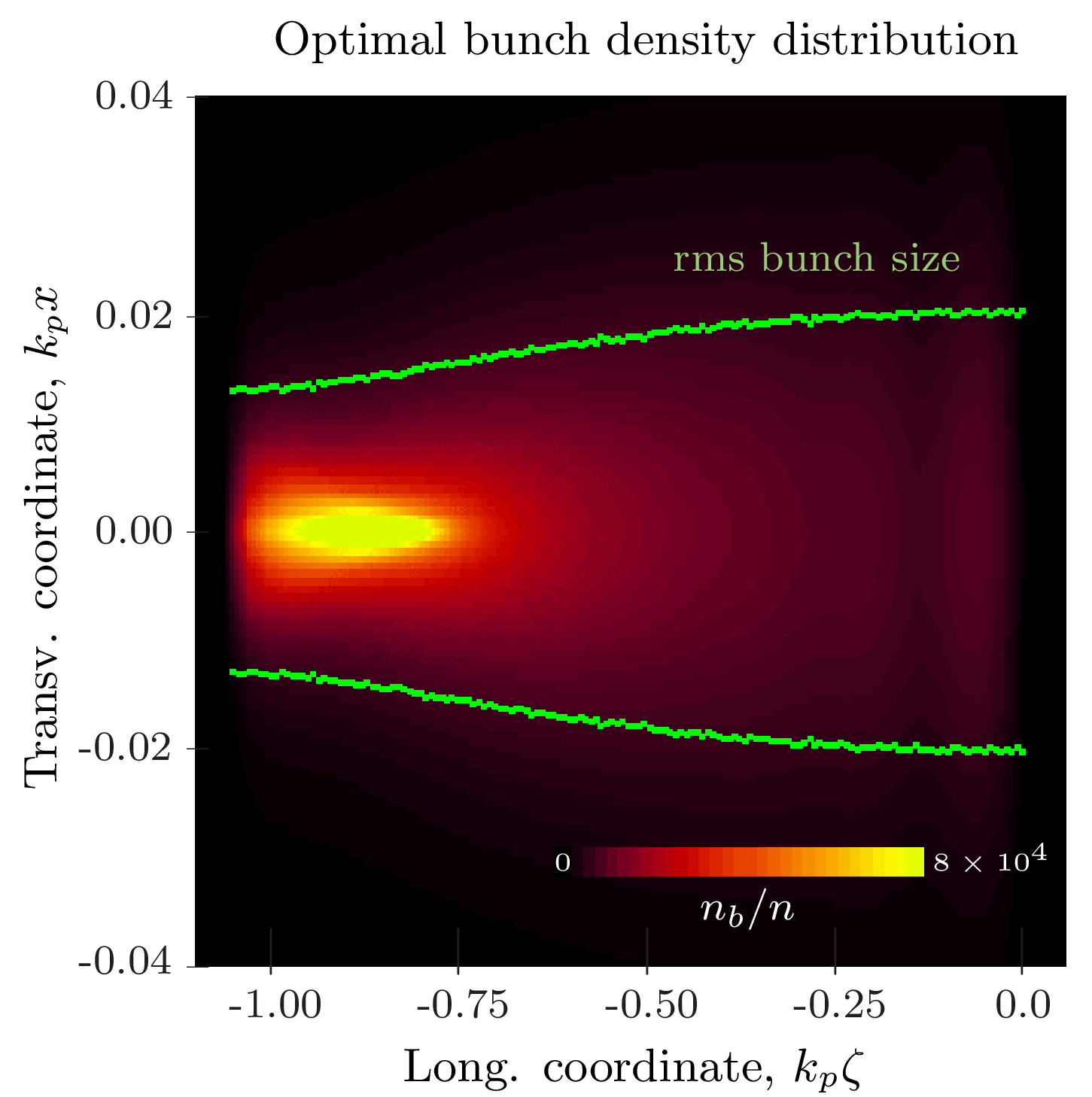}
\caption{\label{fig:beam-dist} 
Two dimensional $(\zeta, x)$ map of the bunch density for the optimal bunch profile that minimizes the energy spread and is an equilibrium solution in the presence of ion motion, suppressing the ion-motion-induced emittance growth. Note that the transverse spatial distribution at any given longitudinal location along the beam is, in general, non-Gaussian.
The bunch energy is 1 GeV, and the normalized slice emittance is 100 nm. The other laser and plasma parameters are the same as in figure~\ref{fig:LPA-stage}.}
\end{figure}

The transverse spot size of the beam may be nonlinearly matched to the ion motion perturbed wakefield eq.~\eqref{eq:wake-ion}, enabling propagation in the nonlinear fields without emittance growth \cite{Benedetti17}. In order to achieve this, the transverse phase-space distribution of the bunch needs to be adjusted slice-by-slice. This ensures a constant slice emittance along the bunch. Note that while the transverse beam momentum distribution remains, slice-by-slice, Gaussian (with slice-dependent rms moments), the transverse spatial distribution required to achieve perfect matching is, in general, non-Gaussian. 
By also longitudinal tailoring the beam current, the accelerating field along the beam can be controlled, resulting in high efficiency and minimizing the energy spread growth.   Figure~\ref{fig:beam-dist} shows the beam distribution that minimizes the energy spread and transverse emittance growth in the ion cavity with ion motion for the LPA stage shown in figure~\ref{fig:LPA-stage}. 
Hence, with longitudinal and transverse bunch shaping, a sufficiently dense beam inducing ion motion may be stably accelerated with beam quality preservation.   This provides a path to high efficiency and quality-preserving acceleration, required for a plasma collider. 
Generation of beams with the desired current profiles could be achieved using, e.g., an emittance exchange beamline \cite{Ha17}, or by means of plasma-based injection schemes \cite{Amorim19}. 
For any given bunch current distribution, transverse tapering of the bunch can be obtained by means of an adiabatic matching procedure \cite{Benedetti21}. 

\subsection{Coulomb scattering}
\label{sec:scatt}

Emittance growth may occur by elastic scattering of the beam electrons against the background ions in the plasma. This  results in a change of the rms divergence of the beam: $d\langle \theta_x^2 \rangle /dt = d\langle \theta_y^2 \rangle /dt  = (1/2) d \langle \theta^2 \rangle /dt$, with \cite{Kirby07,Schroeder13b,Zhao20}
\begin{equation}
\frac{d\langle \theta^2 \rangle }{dt} = \frac{8\pi n r_e^2 c}{\gamma^2} Z_i \ln \Lambda_C,
\end{equation}
where $\ln \Lambda_C = \ln (b_{\rm max}/b_{\rm min})$ is the Coulomb logarithm describing the transverse distance over which the Coulomb interaction is screened.   
The minimum impact parameter $b_{\rm min}$ may be estimated as the Coulombic
radius of the nucleus $b_{\rm min} \sim 1.4 A_i^{1/3}$~fm \cite{Jackson75}.  In the bubble regime, the maximum impact parameter is determined by the electron screening in the sheath of the bubble, $b_{\rm max} \sim R + \lambda_D$, where  $R$ is the bubble radius and $\lambda_D$ is the Debye length. For typical plasma parameters $\lambda_D\ll R \approx 2 \sqrt{a}/k_p \sim \lambda_p$, and the maximum impact parameter may be approximated as $b_{\rm max} \sim \lambda_p$.

\begin{figure}[htbp]
\centering 
\includegraphics[width=.5\textwidth]{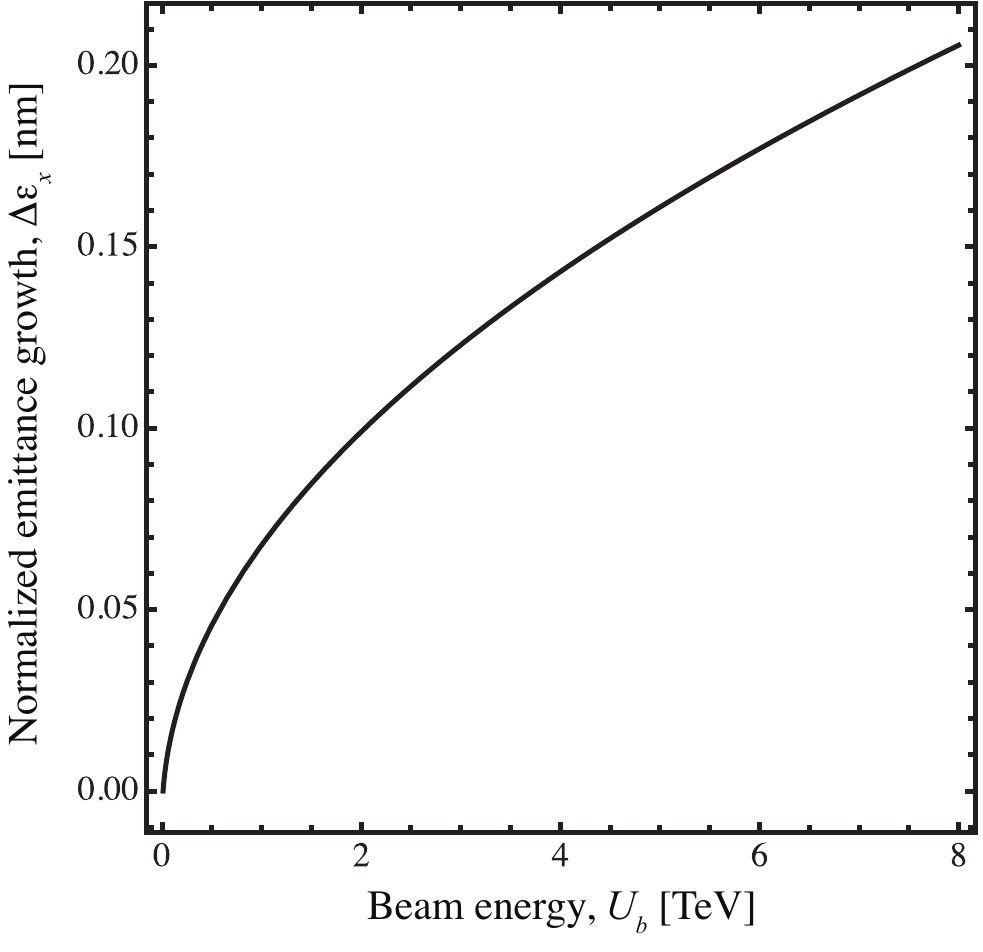}
\caption{\label{fig:scatt} Normalized transverse emittance growth $\Delta \epsilon_x$ owing to Coulomb scattering with Hydrogen ions and with a loaded gradient of $E_z/E_0= 1.8$ and density $5\times10^{17}~\textrm{cm}^{-3}$.
}
\end{figure}

The growth in emittance is \cite{Zhao20} 
\begin{equation}
\frac{d\epsilon_x^2 }{dz} = k_p^2 r_e Z_i \ln \Lambda_C \, \langle x^2 \rangle + 
2 \left[  \langle u_x^2 \rangle \langle x u_x /\gamma \rangle 
- \langle x u_x\rangle \langle u_x^2/\gamma\rangle \right] ,
\end{equation}
where the first term on the right-hand side is due to Coulomb scattering and the second term is the emittance growth due to decoherence associated with energy spread. 
For a monoenergetic, matched beam in an ion cavity, 
\begin{equation}
\frac{d\epsilon_x }{dz} = \frac{k_p r_e}{\sqrt{2\gamma}} Z_i \ln \Lambda_C  .
\end{equation}
Assuming a constant acceleration, $d\gamma/dz = k_p (E_z/E_0)$, the total emittance growth over the length of the accelerator is
 \begin{equation}
\Delta \epsilon_x = \frac{\sqrt{2} r_e Z_i}{(E_z/E_0)}  \ln \Lambda_C \left( \gamma_f^{1/2} - \gamma_i^{1/2}\right) ,
\label{eq:scat}
\end{equation}
where $\gamma_f$ and $\gamma_i$ are the final and initial beam energies, respectively. 
The strong focusing in the ion cavity suppresses the scattering-induced emittance growth.  Note that there is only a very weak dependence on the plasma density,  through the Coulomb logarithm. 
Figure~\ref{fig:scatt} shows the growth in the normalized rms emittance owing to Coulomb scattering with background Hydrogen ions in the plasma.  
Owing to the strong focusing, the emittiance growth is sub-nm for multi-TeV beam energies and, hence, is not the limiting factor for a multi-TeV collider.

\subsection{Betatron radiation}
\label{sec:rad}

As described in sections \ref{sec:hose} and \ref{sec:scatt}, the strong focusing in the plasma creates a dense bunch, providing transverse stability through background ion motion, and mitigates the effects of Coulomb scattering.  However strong transverse focusing will also cause the electrons to emit synchrotron radiation, which can lead to the loss of beam power and beam energy spread growth. 
In general, the power radiated via synchrotron radiation by an electron experiencing a force $F_\perp$ is \cite{Jackson75}
\begin{equation}
P = \frac{2}{3} \frac{e^2 \gamma^2}{m^2c^3} F_\perp^2 .
\end{equation}
If we consider an electron oscillating in the potential of an ion cavity with focusing field given by eq.~\eqref{eq:er}, then the energy loss via radiation, averaged over the betatron period, is 
\begin{equation}
\frac{d\gamma}{dz} = - \frac{ r_e}{12}  \gamma^2 k_p^4 r_\beta^2,
\end{equation}
where $r_\beta$ is the betatron amplitude.  For an accelerating beam, the betatron amplitude decreases adiabatically such that $r_\beta = r_{\beta i} (\gamma_i/\gamma)^{1/4}$, where $r_{\beta i}$ is the betatron amplitude at injection. For collider-relevant beam parameters, the energy loss via radiation is a small perturbation compared to the energy gain from the plasma wave.  Assuming the radiation is a small perturbation (and the accelerating gradient is approximately constant throughout the beam), the energy loss to synchrotron radiation is 
\begin{equation}
\Delta \gamma = - \frac{r_e}{30}  \frac{  k_p^3 \gamma_i^{1/2} \gamma_f^{5/2} 
r_{\beta i}^2 }{ ( E_z/E_0 )} .
\label{eq:dg}
\end{equation}
Averaging over the beam distribution yields, 
\begin{equation}
\langle \Delta \gamma \rangle = - \frac{\sqrt{2}}{30}  r_e \frac{  k_p^2  \gamma_f^{5/2} 
\left( \epsilon_{x} + \epsilon_{y} \right)}{ ( E_z/E_0 )} ,
\end{equation}
where the rms beam size can be expressed in terms of the transverse normalized emittances 
$\langle r_{\beta i}^2 \rangle  = ( \epsilon_{x} + \epsilon_{y} )/(\gamma_i k_{\beta i})$.
For a flat beam, $ \epsilon_{x} \gg  \epsilon_{y} $, the energy loss is determined by the larger of the transverse emittances. 
The total average beam power lost to synchrotron radiation in the plasma is 
$P_{\rm rad} = f N mc^2 \langle  \Delta \gamma \rangle  = P_b \langle  \Delta \gamma \rangle/\gamma$.  The average beam power loss is $P\propto n \gamma_f^{5/2} \epsilon_{x}$.  

\begin{figure}[htbp]
\centering 
\includegraphics[width=.5\textwidth]{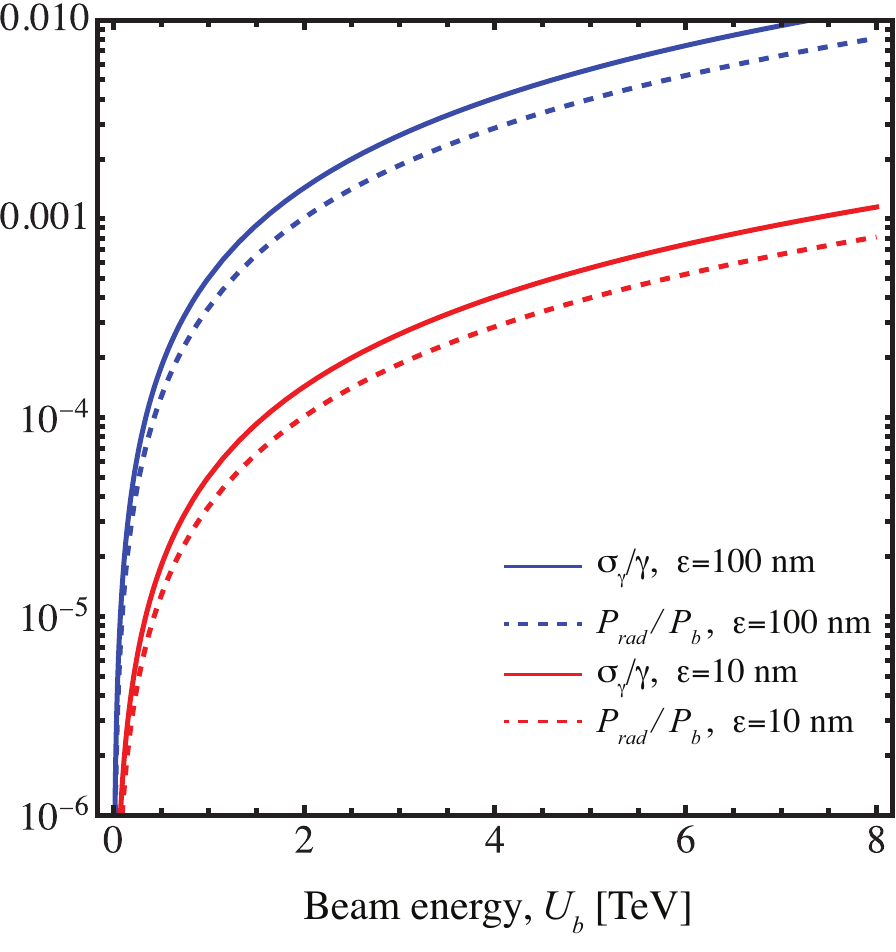}
\caption{\label{fig:synchrad} The growth in the relative energy spread $\sigma_\gamma/\gamma$ (solid curves) and power radiated $P_{\rm rad}/P_b$ (dashed curves) after acceleration in the plasma for a beam with emittance 100 nm (blue curves) and 10 nm (red curves), assuming a density of $n=5 \times 10^{17}$~cm$^{-3}$ and a loaded gradient of $E_z/E_0= 1.8$.}
\end{figure}

On-axis particles will not undergo betatron motion and will not radiate, whereas off-axis particles, with large betaron amplitudes will radiation strongly.   The induced rms beam energy spread is given by $\sigma_\gamma^2 = \langle  (\Delta \gamma - \langle \Delta \gamma \rangle)^2 \rangle$.  Using 
eq.~\eqref{eq:dg}, the growth in relative energy spread owing to synchrotron radiation is 
\begin{equation}
\frac{\sigma_\gamma}{\gamma_f} = \frac{{ r_e}}{15}  \frac{  k_p^2  \gamma_f^{3/2} 
\left( \epsilon_{x} + \epsilon_{y} \right) }{ ( E_z/E_0 )} .
\end{equation}
The synchrotron radiation-induced energy spread in the plasma is 
${\sigma_\gamma}/{\gamma_f} \propto n \gamma_f^{3/2} \epsilon_{x}$.  
Figure~\ref{fig:synchrad} shows the energy spread growth $\sigma_\gamma/\gamma$ (solid curves) and power radiated $P_{\rm rad}/P_b$ (dashed curves) after acceleration in the plasma for a beam with normalized emittance of $\epsilon_{x} + \epsilon_{y} =100$~nm (blue curves) and $\epsilon_{x} + \epsilon_{y} =10$~nm  (red curves), assuming a density of $n=5 \times 10^{17}$~cm$^{-3}$ and a loaded gradient of $E_z/E_0= 1.8$.  Provided a low emittance, matched beam is accelerated in the plasma, the energy spread growth is a fraction of a percent for multi-TeV colliders.

\section{Conclusions and discussion}

In this work we have discussed some design considerations for a laser-plasma accelerator-based linear collider.  LPAs offer ultra-high gradients, $E_z \sim E_0$, and ultra-short bunches, $\sigma_z \sim k_p^{-1}$.  These two characteristics offer the potential for compact linacs and reduced power requirements, as described in section \ref{sec:general}.  An example of an LPA stage in the nonlinear bubble regime was presented in section \ref{sec:LPA}.    Several of the most important beam dynamics challenges unique to plasma accelerators were considered. 
Transverse stability (i.e., beam hosing) has been identified as a potential limit to the application of plasma accelerators.   Here we show that the ion motion that naturally occurs for dense beams (i.e., low emittance and high energy) will severely damp the beam hosing.  By injecting a beam with a transverse distribution that is matched to the wakefield with ion motion, an electron bunch may be accelerated stably with high efficiency, without energy spread or emittance growth.  We have also considered Coulomb scattering by the background ions in section \ref{sec:scatt}.  The strong focusing in the plasma strongly suppresses emittance growth by scattering.  In the nonlinear bubble regime scattering results in sub-nm normalized emittance growth for multi-TeV-class plasma linacs.  The strong focusing also results in betatron radiation emission.  Betatron emission will result in beam power loss and energy spread growth. For sub-100-nm emittances, the  beam power loss and energy spread growth will be sub-percent for multi-TeV-class plasma linacs.

Efficient, beam-quality-preserving positron acceleration is a challenge for plasma accelerators.  In the nonlinear bubble regime the positron beam will be defocused in the ion cavity.  Hollow plasma channels \cite{Gessner16,Schroeder16}, to remove the defocusing background ions, have been considered for positron acceleration; however hollow plasma channels are subject to severe transverse instabilities \cite{Schroeder99a}. Plasma columns \cite{Diederichs19}, which generate an extended accelerating and focusing region for positrons in the wakefield, have recently been proposed and look promising, and further analysis is in progress \cite{Diederichs20}.  High efficiency positron acceleration, comparable to electron acceleration, remains a challenge.  Owing to the challenges of positron acceleration in plasma, a $\gamma\gamma$ collider may be considered. In a $\gamma\gamma$ collider, two electron beam linacs would be employed and Compton backscattering would be used near the IP to generate colliding energetic photon beams.  Photon collisions can access many of the lepton interactions available in an $e^{+}e^{-}$ collider, and from a collider design viewpoint, a $\gamma\gamma$ collider eliminates the need for positron beam acceleration in plasma, and removes the beamstrahlung and beam-beam constraints described in section \ref{sec:general}.  Photon-electron collisions could also be considered, with a potential higher collision energy reach, compared to $\gamma\gamma$. Without beamstrahlung constraints, the power required to achieve a luminosity could be reduced by operating at higher electron charge per bunch.  The challenge for a $\gamma\gamma$ (or $\gamma e^{-}$) collider, is developing the Compton scattering source at the appropriate wavelength and average power.  To avoid pair creation during scattering \cite{Telnov95}, the incident photon wavelength must be longer than $\lambda [\mu\textrm{m}] > 3.93 U_b$[TeV], and at this wavelength yields a peak scattered photon energy of $\hbar \omega \simeq 0.82 U_b$[TeV]. For example, a $7.2~\mu$m wavelength laser would be required to scatter off a 1.8~TeV electron beam to produce 1.5~TeV photons. With a conversion efficiency (scattered photons per electron) of 0.65, approximately 0.7~TW of laser power per pulse is required. Development of high-average-power laser sources for scattering in the mid-IR regime presents a technical challenge to development of $\gamma\gamma$ colliders operating at multi-TeV energies.

A multi-TeV lepton linear collider has the potential to open new possibilities for high energy particle physics research.  Plasma accelerators enable collider size and power reductions, compared to accelerators based on conventional RF technology, although significant research and development is required to bring plasma accelerator technology to sufficient maturity to be considered for the next generation collider.  The beam dynamics considerations in this work indicate that multi-TeV plasma linacs are achievable. With the rapid development of plasma accelerators over the past decade, the community has begun to examine the integration of collider subsystems compatible with plasma accelerators, with the goal of completing an integrated design study of a collider based on plasma accelerators over the next several years. 


\acknowledgments
This work was supported by the Director, Office of Science, Office of High Energy Physics, of the U.S. Department of Energy under Contract No.\ DE-AC02-05CH11231 and used the computational facilities at the National Energy Research Scientific Computing Center (NERSC).

\providecommand{\href}[2]{#2}\begingroup\raggedright\endgroup


\begin{thebibliography}{10}

\bibitem{CLIC-CDR}
\emph{{CLIC} {Conceptual Design Report}},  Tech. Rep. CERN-2012-007, CERN
  (2012).

\bibitem{Esarey09}
E.~Esarey, C.B.~Schroeder and W.P.~Leemans, \emph{Physics of laser-driven
  plasma-based electron accelerators}, {\emph{Rev. Mod. Phys.} {\bfseries 81}
  (2009) 1229}.

\bibitem{Hooker13}
S.M.~Hooker, \emph{Developments in laser-driven plasma accelerators},
  {\emph{Nature Photonics} {\bfseries 7} (2013) 775}.

\bibitem{Gonsalves19}
A.J.~Gonsalves, K.~Nakamura, J.~Daniels, C.~Benedetti, C.~Pieronek,
  T.C.H.~de~Raadt et~al., \emph{Petawatt laser guiding and electron beam
  acceleration to {8 GeV} in a laser-heated capillary discharge waveguide},
  {\emph{Phys. Rev. Lett.} {\bfseries 122} (2019) 084801}.

\bibitem{Schroeder10b}
C.B.~Schroeder, E.~Esarey, C.G.R.~Geddes, C.~Benedetti and W.P.~Leemans,
  \emph{Physics considerations for laser-plasma linear colliders}, {\emph{Phys.
  Rev. ST Accel. Beams} {\bfseries 13} (2010) 101301}.

\bibitem{Schroeder16}
C.~Schroeder, C.~Benedetti, E.~Esarey and W.~Leemans, \emph{Laser-plasma-based
  linear collider using hollow plasma channels}, {\emph{Nucl. Instrum. Methods
  Phys. Res. A} {\bfseries 829} (2016) 113}.

\bibitem{ANAR17}
B.~Cros and P.~Muggli, eds., \emph{Towards a Proposal for an Advanced Linear
  Collider, Report on the Advanced and Novel Accelerators for High Energy
  Physics Roadmap Workshop}, no.~CH-1211, CERN, 2017.

\bibitem{Oide88}
K.~Oide, \emph{Synchrotron-radiation limit on the focusing of electron beams},
  {\emph{Phys. Rev. Lett.} {\bfseries 61} (1988) 1713}.

\bibitem{YokoyaChen92}
K.~Yokoya and P.~Chen, \emph{Beam-beam phenomena in linear colliders},  in
  \emph{Frontiers of Particle Beams: Intensity Limitations}, M.~Dienes,
  M.~Month and S.~Turner, eds., vol.~400 of \emph{Lecture Notes in Physics},
  (Berlin), pp.~415--445, Springer-Verlag, 1992.

\bibitem{Dugan94}
G.~Dugan, \emph{Advanced accelerator system requirements for future linear
  colliders}, {\emph{AIP Conference Proceedings} {\bfseries 737} (2004) 29}.

\bibitem{Diederichs19}
S.~Diederichs, T.J.~Mehrling, C.~Benedetti, C.B.~Schroeder, A.~Knetsch,
  E.~Esarey et~al., \emph{Positron transport and acceleration in beam-driven
  plasma wakefield accelerators using plasma columns}, {\emph{Phys. Rev. Accel.
  Beams} {\bfseries 22} (2019) 081301}.

\bibitem{Lu07}
W.~Lu, M.~Tzoufras, C.~Joshi, F.S.~Tsung, W.B.~Mori, J.~Vieira et~al.,
  \emph{Generating multi-{GeV} electron bunches using single stage laser
  wakefield acceleration in a {3D} nonlinear regime}, {\emph{Phys. Rev. ST
  Accel. Beams} {\bfseries 10} (2007) 061301}.

\bibitem{Benedetti13}
C.~Benedetti, C.B.~Schroeder, E.~Esarey, F.~Rossi and W.P.~Leemans,
  \emph{Numerical investigation of electron self-injection in the nonlinear
  bubble regime}, {\emph{Phys. Plasmas} {\bfseries 20} (2013) 103108}.

\bibitem{INFERNO1}
C.~Benedetti, C.B.~Schroeder, E.~Esarey, C.G.R.~Geddes and W.P.~Leemans,
  \emph{Efficient modeling of laser‐plasma accelerators with {INF\&RNO}},
  {\emph{AIP conference proceedings} {\bfseries 1299} (2010) 250}.

\bibitem{INFERNO2}
C.~Benedetti, C.B.~Schroeder, E.~Esarey, C.G.R.~Geddes and W.P.~Leemans,
  \emph{An accurate and efficient laser-envelope solver for the modeling of
  laser-plasma accelerators}, {\emph{Plasma Phys. Control. Fusion} {\bfseries
  60} (2018) 014002}.

\bibitem{steinke16}
S.~Steinke, J.~van Tilborg, C.~Benedetti, C.G.R.~Geddes, C.B.~Schroeder,
  J.~Daniels et~al., \emph{Multistage coupling of independent laser-plasma
  accelerators}, {\emph{Nature} {\bfseries 530} (2016) 190}.

\bibitem{Zhou15}
T.~Zhou, J.~Ruppe, C.~Zhu, I.-N.~Hu, J.~Nees and A.~Galvanauskas,
  \emph{Coherent pulse stacking amplification using low-finesse gires-tournois
  interferometers}, {\emph{Opt. Express} {\bfseries 23} (2015) 7442}.

\bibitem{Stark21}
H.~Stark, J.~Buldt, M.~M\"{u}ller, A.~Klenke and J.~Limpert, \emph{{1 kW, 10
  mJ, 120 fs} coherently combined fiber {CPA} laser system}, {\emph{Opt. Lett.}
  {\bfseries 46} (2021) 969}.

\bibitem{Tamer21}
I.~Tamer, B.A.~Reagan, T.~Galvin, J.~Galbraith, E.~Sistrunk, A.~Church et~al.,
  \emph{Demonstration of a compact, multi-joule, diode-pumped {Tm:YLF} laser},
  {\emph{Opt. Lett.} {\bfseries 46} (2021) 5096}.

\bibitem{Whittum91}
D.H.~Whittum, W.M.~Sharp, S.S.~Yu, M.~Lampe and G.~Joyce, \emph{Electron-hose
  instability in the ion-focused regime}, {\emph{Phys. Rev. Lett.} {\bfseries
  67} (1991) 991}.

\bibitem{Huang07}
C.~Huang, W.~Lu, M.~Zhou, C.E.~Clayton, C.~Joshi, W.B.~Mori et~al.,
  \emph{Hosing instability in the blow-out regime for plasma-wakefield
  acceleration}, {\emph{Phys. Rev. Lett.} {\bfseries 99} (2007) 255001}.

\bibitem{Mehrling18b}
T.J.~Mehrling, C.~Benedetti, C.B.~Schroeder, A.~Martinez de~la Ossa,
  J.~Osterhoff, E.~Esarey et~al., \emph{Accurate modeling of the hose
  instability in plasma wakefield accelerators}, {\emph{Phys. Plasmas}
  {\bfseries 25} (2018) 056703}.

\bibitem{Lebedev17}
V.~Lebedev, A.~Burov and S.~Nagaitsev, \emph{Efficiency versus instability in
  plasma accelerators}, {\emph{Phys. Rev. Accel. Beams} {\bfseries 20} (2017)
  121301}.

\bibitem{YYLau89}
Y.Y.~Lau, \emph{Classification of beam breakup instabilities in linear
  accelerators}, {\emph{Phys. Rev. Lett.} {\bfseries 63} (1989) 1141}.

\bibitem{Lehe17}
R.~Lehe, C.B.~Schroeder, J.-L.~Vay, E.~Esarey and W.P.~Leemans,
  \emph{Saturation of the hosing instability in quasilinear plasma
  accelerators}, {\emph{Phys. Rev. Lett.} {\bfseries 119} (2017) 244801}.

\bibitem{Rosenzweig05}
J.B.~Rosenzweig, A.M.~Cook, A.~Scott, M.C.~Thompson and R.B.~Yoder,
  \emph{Effects of ion motion in intense beam-driven plasma wakefield
  accelerators}, {\emph{Phys. Rev. Lett.} {\bfseries 95} (2005) 195002}.

\bibitem{Benedetti17}
C.~Benedetti, C.B.~Schroeder, E.~Esarey and W.P.~Leemans, \emph{Emittance
  preservation in plasma-based accelerators with ion motion}, {\emph{Phys. Rev.
  Accel. Beams} {\bfseries 20} (2017) 111301}.

\bibitem{An17}
W.~An, W.~Lu, C.~Huang, X.~Xu, M.J.~Hogan, C.~Joshi et~al., \emph{Ion motion
  induced emittance growth of matched electron beams in plasma wakefields},
  {\emph{Phys. Rev. Lett.} {\bfseries 118} (2017) 244801}.

\bibitem{Ha17}
G.~Ha, M.H.~Cho, W.~Namkung, J.G.~Power, D.S.~Doran, E.E.~Wisniewski et~al.,
  \emph{Precision control of the electron longitudinal bunch shape using an
  emittance-exchange beam line}, {\emph{Phys. Rev. Lett.} {\bfseries 118}
  (2017) 104801}.

\bibitem{Amorim19}
L.D.~Amorim, N.~Vafaei-Najafabadi, C.~Emma, C.I.~Clarke, S.Z.~Green, D.~Storey
  et~al., \emph{Shaping trailing beams for beam loading via
  beam-induced-ionization injection at facet}, {\emph{Phys. Rev. Accel. Beams}
  {\bfseries 22} (2019) 111303}.

\bibitem{Benedetti21}
C.~Benedetti, T.J.~Mehrling, C.B.~Schroeder, C.G.R.~Geddes and E.~Esarey,
  \emph{Adiabatic matching of particle bunches in a plasma-based accelerator in
  the presence of ion motion}, {\emph{Phys. Plasmas} {\bfseries 28} (2021)
  053102}.

\bibitem{Kirby07}
N.~Kirby, M.~Berry, I.~Blumenfeld, M.J.~Hogan, R.~Ischebeck and R.~Siemann,
  \emph{Emittance growth from multiple coulomb scattering in a plasma wakefield
  accelerator},  in \emph{Proceedings of PAC07}, (www.jacow.org),
  pp.~3097--3099, JACoW, 2007.

\bibitem{Schroeder13b}
C.B.~Schroeder, E.~Esarey, C.~Benedetti and W.P.~Leemans, \emph{Control of
  focusing forces and emittances in plasma-based accelerators using near-hollow
  plasma channels}, {\emph{Phys. Plasmas} {\bfseries 20} (2013) 080701}.

\bibitem{Zhao20}
Y.~Zhao, R.~Lehe, A.~Myers, M.~Th{\'e}venet, A.~Huebl, C.B.~Schroeder et~al.,
  \emph{Modeling of emittance growth due to coulomb collisions in plasma-based
  accelerators}, {\emph{Phys. Plasmas} {\bfseries 27} (2020) 113105}.

\bibitem{Jackson75}
J.D.~Jackson, \emph{Classical Electrodynamics}, Wiley, New York, 2nd~ed.
  (1975).

\bibitem{Gessner16}
S.~Gessner, E.~Adli, J.M.~Allen, W.~An, C.I.~Clarke, C.E.~Clayton et~al.,
  \emph{Demonstration of a positron beam-driven hollow channel plasma wakefield
  accelerator}, {\emph{Nature Communications} {\bfseries 7} (2016) 11785}.

\bibitem{Schroeder99a}
C.B.~Schroeder, D.H.~Whittum and J.S.~Wurtele, \emph{Multimode analysis of the
  hollow plasma channel accelerator}, {\emph{Phys. Rev. Lett.} {\bfseries 82}
  (1999) 1177}.

\bibitem{Diederichs20}
S.~Diederichs, C.~Benedetti, E.~Esarey, J.~Osterhoff and C.B.~Schroeder,
  \emph{High-quality positron acceleration in beam-driven plasma accelerators},
  {\emph{Phys. Rev. Accel. Beams} {\bfseries 23} (2020) 121301}.

\bibitem{Telnov95}
V.~Telnov, \emph{Principles of photon colliders}, {\emph{Nucl. Instrum. Methods
  Phys. Res. A} {\bfseries 355} (1995) 3}.

\end{thebibliography}

\end{document}